\documentclass[conference,letter]{IEEEtran}
\usepackage[letterpaper, left=0.625in, right=0.625in, bottom=1.1in, top=2.0cm]{geometry}

\usepackage{graphicx,calc}
\usepackage{paralist}
\usepackage{balance}
\usepackage{cite}
\usepackage{booktabs}
\usepackage[hidelinks]{hyperref}
\usepackage{glossaries}
\glsdisablehyper
\usepackage{subcaption}
\usepackage{eso-pic}
\begin{document}
\AddToShipoutPictureBG*{\AtPageUpperLeft{\put(\LenToUnit{0.625in},\LenToUnit{-0.6cm}){\parbox[t]{\dimexpr\paperwidth-1.25in\relax}{\centering\scriptsize
\copyright~2026 IEEE. Personal use of this material is permitted. Permission from IEEE must be obtained for all other uses, in any current or future media, including reprinting/republishing this material for advertising or promotional purposes, creating new collective works, for resale or redistribution to servers or lists, or reuse of any copyrighted component of this work in other works.\\
Published in: 2026 IEEE International Conference on Cyber Security and Resilience (CSR), Lisbon, Portugal, 2026, pp.~218--223. DOI: \href{https://doi.org/10.1109/CSR69374.2026.11760099}{10.1109/CSR69374.2026.11760099}}}}}

\title{Autonomous Cyber Operations Gym for Training and Evaluating AI Models for Defence}

        \author{
  \IEEEauthorblockN{
         Tobias Hürten$^{1}$, Johannes F. Loevenich$^{1,2}$, Florian Spelter$^{1}$, Erik Adler$^{1,3}$, \\ Laurin Holz$^{1,4}$, Maxime Schwarzer$^{1,3}$, Thies Moehlenhof$^{1}$, and Roberto Rigolin F. Lopes$^{1}$}  
  \IEEEauthorblockA{
        $^{1}$CortAIx Labs, Thales Deutschland, Ditzingen, Germany \\
        $^{2}$Department of Mathematics/Computer Science, University of Osnabrück, Osnabrück, Germany \\        
        $^{3}$Department of Informatics, Karlsruhe Institute of Technology (KIT), Karlsruhe, Germany \\
        $^{4}$Department of Computer Science, University of Ulm, Ulm, Germany \\
        Email: \{tobias.huerten, johannes.loevenich, florian.spelter, erik.adler, laurin.holz, \\
        maxime.schwarzer, thies.moehlenhof, roberto.rigolin\}@thalesgroup.com
    }
}

\maketitle

        \newacronym{ACO}{ACO}{Automated Cyber Operation}
\newacronym{AI}{AI}{Artificial Intelligence}
\newacronym{API}{API}{Application Programming Interface}
\newacronym{ACD}{ACD}{Autonomous Cyber Defence}
\newacronym{ABM}{ABM}{Agent-Based Modelling}
\newacronym{BGP}{BGP}{Border Gateway Protocol}
\newacronym{C2}{C2}{Command and Control}
\newacronym{CR}{CR}{Cyber Range}
\newacronym{CM}{CM}{Cognitive Modelling}
\newacronym{DPI}{DPI}{Deep Packet Inspection}
\newacronym[plural=DNNs]{DNN}{DNN}{Deep Neural Network}
\newacronym{DRL}{DRL}{Deep Reinforcement Learning}
\newacronym{DNS}{DNS}{Domain Name System}
\newacronym{FRR}{FRR}{Free Range Routing}
\newacronym{FP}{FP}{False Positive}
\newacronym{FN}{FN}{False Negative}
\newacronym{GenAI}{GenAI}{Generative AI}
\newacronym{HTTP}{HTTP}{Hypertext Transfer Protocol}
\newacronym[plural=IDSs]{IDS}{IDS}{Intrusion Detection System}
\newacronym{IP}{IP}{Internet Protocol}
\newacronym{ISP}{ISP}{Internet Service Provider}
\newacronym[plural=LLMs]{LLM}{LLM}{Large Language Model}
\newacronym{ML}{ML}{Machine Learning}
\newacronym{NATO}{NATO}{North Atlantic Treaty Organization}
\newacronym{NPC}{NPC}{Non-Player Character}
\newacronym{OSPF}{OSPF}{Open Shortest Path First}
\newacronym{PCN}{PCN}{Protected Core Networking}
\newacronym{RL}{RL}{Reinforcement Learning}
\newacronym{RCE}{RCE}{Remote Code Execution}
\newacronym{ReLU}{ReLU}{Rectified Linear Unit}
\newacronym{SDD}{SDD}{Software-Defined Defence}
\newacronym{SDN}{SDN}{Software Defined Networking}
\newacronym{SSTI}{SSTI}{Server Side Template Injection}
\newacronym{SMB}{SMB}{Server Message Block}
\newacronym{SM}{SM}{Statistical Modelling}
\newacronym{TCP}{TCP}{Transmission Control Protocol}
\newacronym{TC}{TC}{Traffic Control}
\newacronym{TP}{TP}{True Positive}
\newacronym{UDP}{UDP}{User Datagram Protocol}

        \begin{abstract}
This paper proposes an Autonomous Cyber Operations (ACO) gym, a unified framework integrating high-fidelity network emulation, realistic user behaviour modelling for benign network traffic and adversarial profiles, and the generation of cybersecurity scenarios through open toolkits. By supporting live and deferred feature extraction at the packet and flow levels, the gym facilitates the creation of labelled synthetic datasets for training Deep Neural Network (DNN)-based Intrusion Detection Systems (IDSs), as well as developing and evaluating modular components for autonomous defence agents through emulated interactions. While evaluation of a DNN classifier within the gym shows high in-domain performance, achieving near-perfect recall on adversarial network flows, our experiments reveal a significant sensitivity to domain shift, showing that even minor changes in network composition, such as variations in traffic protocol distribution, can lead to a catastrophic collapse in model transferability and detection performance.
\end{abstract}


        \section{Introduction}
\label{sec:introduction}
The fundamental reshaping of the cybersecurity landscape by the increasing volume, speed, and sophistication of cyberattacks motivates a shift toward \gls{ACD}, where \gls{AI} systems, such as \gls{DRL}~\cite{sutton1998reinforcement} or \glspl{LLM}~\cite{loevenich2024journal, Hurten:MILCOM2024}, support detection and timely response in complex environments~\cite{vyas2023automated}.
To this end, this work focuses on the design of an extendable environment that supports key \gls{SDD} principles to emulate realistic network topologies and user behaviour models, as well as the evaluation of an \gls{IDS} under protocol-distribution shift.

One of the challenges of robust \gls{ACD} agents is the lack of realistic and rigorous training/testing environments~\cite{vyas2023automated}. Effective \gls{ACD} solutions, whether \gls{DNN}-based \glspl{IDS} or \gls{DRL}-based autonomous defence agents, require vast quantities of high-fidelity, labelled network traffic data (either as aggregated datasets or through live interactions) and a controlled, reproducible environment. Existing platforms often fail to balance high fidelity with scalability, restrict functionality to solely offensive or defensive research, or lack realistic benign user traffic~\cite{velazquez2025cyber}. Accordingly, this work focuses on high-fidelity environment design and baseline \gls{IDS} evaluation as a step toward enabling full \gls{ACD} systems, addressing fidelity, data generation, and configurability gaps.

The main contributions are:
\begin{itemize}
    \item A modular \gls{ACO} gym integrating emulation, stochastic user behaviour modelling, and synthetic, labelled data generation.
    \item A dual-pipeline framework for both live and deferred network feature extraction.
    \item An empirical analysis of model robustness and cross-domain generalization of an \gls{IDS} under protocol-distribution shift.
\end{itemize}

The remainder of the paper is structured as follows: Section~\ref{sec:related_work} reviews related work regarding \gls{ACD}, the role of cyber ranges in \gls{ACO} gyms, and behavioural modelling. 
Section~\ref{sec:methodology} defines the methodology, including the system architecture, user simulation framework, scenario topology, and threat landscape. Next, Section~\ref{sec:results} outlines the experimental setup, provides a statistical analysis of the generated synthetic datasets, and evaluates a \gls{DNN}-based \gls{IDS} with respect to in-domain performance and transferability under cross-evaluation. Finally, Section~\ref{sec:conclusion} concludes the paper and outlines directions for future work.

        \section{Related Work}
\label{sec:related_work}

The paradigm shift from conventional automation to \gls{ACD} is necessitated by the increasing volume and complexity of \gls{AI}-powered cyber threats~\cite{loevenich2024journal}. Unlike traditional security orchestration, which relies on static, fixed-purpose, rule-based logic, \gls{ACD} emphasises autonomous, adaptive, sequential decision-making agents capable of detecting and mitigating complex attacks without direct human tasking~\cite{vyas2023automated}. This transition addresses the limitations of manual intervention and static response systems in the face of modern threat scales~\cite{velazquez2025cyber}. To ensure reliability, particularly within critical infrastructure, current literature identifies several high-level requirements for trustworthy \gls{ACD}: adaptability to evolving landscapes, auditability for transparent and explainable decision-making, directability for human-in-the-loop intervention, observability for situational awareness, security against adversarial \gls{ML} attacks, and transferability across heterogeneous real-world environments~\cite{lohn2023autonomous}. Closely related is the concept of \acrfull{ACO}, which involves the concurrent deployment of automated red and blue agents within a networked environment. \gls{ACO} serves as a critical framework for simulating the competitive ``cyber arms race'', particularly where network scale or communication constraints render human-led operations infeasible~\cite{vyas2023automated}.

\gls{ACO} gyms serve as a critical framework to bridge the ``deployment gap'' and ensure that \gls{ACD} solutions meet the rigorous standards of critical infrastructure, such as \gls{NATO} \gls{PCN}~\cite{loevenich2024journal, vyas2023automated}. These gyms provide controlled experimental environments built upon \glspl{CR}: isolated, interactive platforms designed to replicate critical, enterprise, and civilian network infrastructures for cybersecurity research and education~\cite{davis2013survey, lillemets2025systematic}. \gls{CR} architectures are broadly categorised into simulated and emulated environments. 
While existing platforms like CybORG~\cite{standen2021cyborg} provide robust \gls{RL} environments, they often rely on abstract, simulated \glspl{CR} for their high scalability and cost-effectiveness~\cite{kavak2021simulation, molina2021network, vyas2023automated}.
In contrast, our approach utilises an emulated \gls{CR} executing authentic software on virtualised hardware to prioritise granular application-level behaviours and evaluation under protocol-distribution shift, critical factors for real-world \gls{IDS} reliability.
Although more resource-intensive and complex to deploy, emulated \glspl{CR} are preferred for high-stakes experimentation, as their increased realism significantly narrows the deployment gap compared to pure simulation~\cite{kavak2021simulation, molina2021network, vyas2023automated}.

High-fidelity \glspl{CR} within the context of \gls{ACO} environments necessitate the generation of realistic, non-deterministic user activity to establish background noise as operational baselines, simulating human-centric failure points, such as phishing, as well as sophisticated adversarial behaviour like non-deterministic multi-stage attack chains~\cite{loevenich2025automating, lohn2023autonomous, sharafaldin2018toward, velazquez2025cyber}. Current literature identifies several modelling paradigms to address these requirements. \gls{SM} replicates technical traffic properties like inter-packet times and payload sizes as random variables but is constrained by its reliance on pre-existing empirical data~\cite{botta2012tool}. Conversely, \gls{CM} captures behavioural intent and psychological factors to predict decision-making biases~\cite{veksler2020cognitive}, while \gls{ABM} simulates emergent macro-level interactions, such as malware propagation across diverse topologies~\cite{vestad2024survey}. The emergence of \gls{GenAI} further enhances these capabilities by synthesising complex, non-repetitive communication and activity patterns, thereby bypassing the privacy and availability constraints inherent in real-world datasets~\cite{wang2025user}. Integrating these methodologies, frameworks like GHOSTS utilise autonomous \glspl{NPC} to achieve role-related objectives through the actuation of real software applications rather than mere simulation. This approach ensures that generated network artifacts, including headers, timing, and application-layer details, maintain the high fidelity required for effective cyber warfare training and the evaluation of autonomous defence agents~\cite{updyke2018ghosts}.
        
        \section{Methodology} \label{sec:methodology}

\subsection{Gym Architecture}
\begin{figure}[!t]
    \centering
        \includegraphics[width=1\columnwidth]{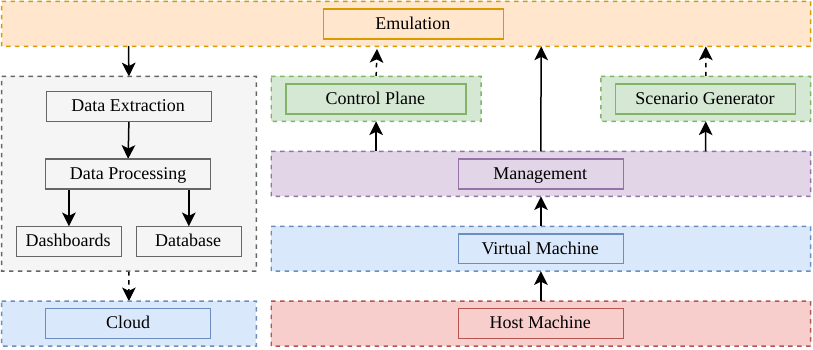}    
    \caption{Architecture of the \acrshort{ACO} gym.}
    \label{fig:architecture}
    \vspace{-1.0\baselineskip}
\end{figure}

Adhering to \gls{SDD} principles, the \gls{ACO} gym's modular architecture decouples hardware, software, and data, providing an extensible framework for high-fidelity emulation of civilian and critical infrastructure through user-defined scenarios.

The system architecture is structured into multiple layers, as illustrated in Fig.~\ref{fig:architecture}. The core component is a network emulation built upon Kathará~\cite{bonofiglio2018kathara} that leverages Docker for resource-efficient containerization. This setup allows for the flexible configuration and instantiation of numerous network nodes, acting as various \gls{SDN} components, such as switches, routers, controllers, hosts, and servers. The entire emulation environment is virtualised to isolate system and hardware dependencies, guaranteeing compatibility, flexibility, and configurability. This encapsulation permits the parallel execution of multiple scenario emulations without mutual interference, thereby facilitating scalable data generation.

A management system serves as the control plane, offering an interface for managing emulation lifecycles and enabling manual control of individual emulated nodes. This system provides the access point for future \gls{DRL} agents to perform decentralised administrative network actions, which effectively decouples the development and training of \gls{ACD} systems from the emulation itself.

\subsection{User Simulation}
To simulate high-fidelity network traffic for training \gls{ACD} systems, we developed \emph{Droids}, a lightweight, modular task-scheduling framework inspired by works such as GHOSTS~\cite{updyke2018ghosts}. \emph{Droids} employs a hierarchical, YAML-based configuration to orchestrate complex automation timelines through a taxonomy of three distinct job classes: \emph{Composite Jobs} provide flow control and concurrency, \emph{Stochastic Jobs} introduce non-deterministic variability to mimic the unpredictability of human actors, and \emph{Primitive Jobs} execute the low-level actions required to generate authentic system and network telemetry.

In practice, this modularity enables reproducible yet stochastic modelling of heterogeneous agents via YAML-parametrised profiles (e.g., activity frequency, \gls{SMB}/\gls{HTTP} protocol mix, and timing jitter). In our case study, benign users are simulated using layered, concurrent workflows that incorporate web crawling and probabilistic file server interactions, while adversarial agents employ the same stochastic building blocks to execute ``low-and-slow'' exploits. By integrating these varied job types, \emph{Droids} generates the concurrent \gls{HTTP}, \gls{SMB}, and system-level traffic with stochastic timing and mixed benign/adversarial activity to quantify environmental fidelity.

\subsection{Network Topology}
\begin{figure}[!htb]
    \centering
        \includegraphics[width=1\columnwidth]{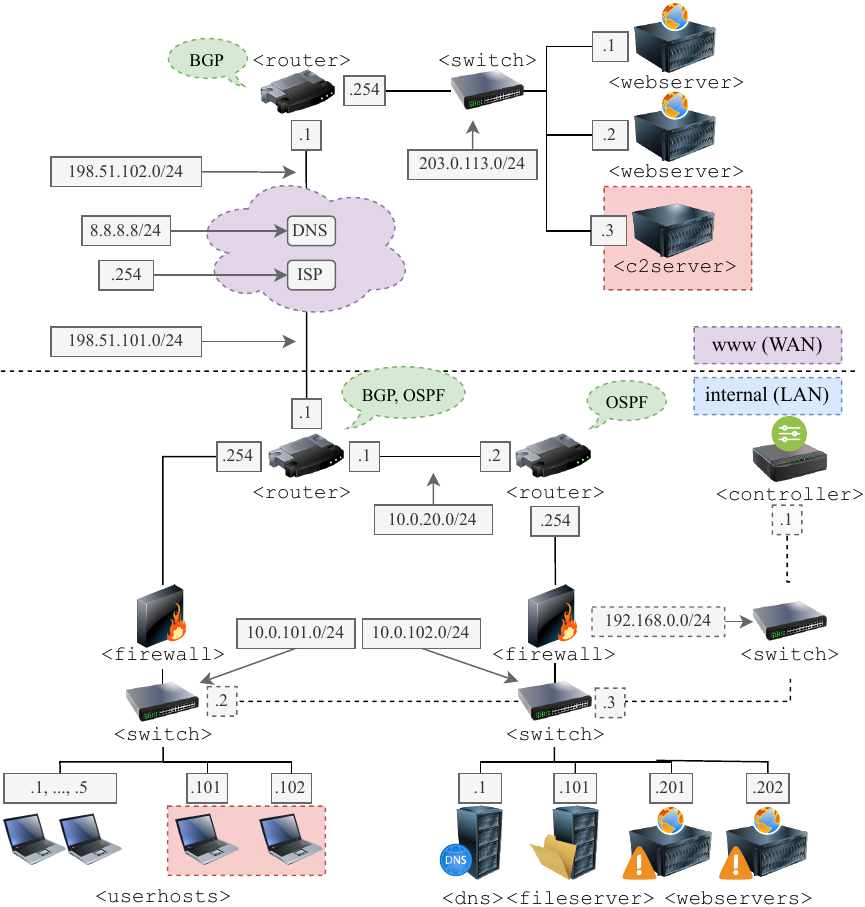} 
    \caption{Exemplary network topology of an enterprise cybersecurity scenario.}
    \label{fig:scenario}
    \vspace{-1.0\baselineskip}
\end{figure}

The emulated enterprise environment, illustrated in Fig.~\ref{fig:scenario}, comprises distinct WAN and LAN segments. The internal LAN is partitioned into two subnets: an outer subnet (\texttt{10.0.101.0/24}) hosting five authorised users and two pre-compromised hosts, and an isolated inner subnet (\texttt{10.0.102.0/24}) providing critical services such as a \gls{DNS} server, internal web services, and a file server (Samba). The external WAN includes an adversarial \gls{C2} server, standard internet services (\gls{ISP}, \gls{DNS}), and two public web servers.

Routing and control follow the \gls{SDN} paradigm, utilising a Floodlight controller and OpenFlow via Open vSwitch. Internal routing employs the \gls{OSPF} protocol, while the \gls{BGP} manages external traffic, intentionally excluded from the inner subnet to ensure isolation. Routers are implemented via the \gls{FRR} suite, with firewalls strictly blocking external traffic to the internal segment. Realistic traffic conditions and entity behaviours (authorized and adversarial) are generated using \gls{TC} and the \emph{Droids} framework, respectively. The scenario definition follows Kathará's configuration schema with general connectivity, as well as host-based start-up and configuration specifications.

\subsection{Threat Analysis}
The illustrative scenario models an assumed breach scenario: a sophisticated internal threat originating from a pre-compromised host in the outer LAN subnet. The primary attack vector targets a \gls{SSTI} vulnerability in the internal Flask web servers. Beyond basic reconnaissance and Samba-based data theft, the adversary executes a complex, multi-stage exfiltration chain:
\begin{inparaenum}[(1)]
\item the adversary first maps the network to identify connectivity between the compromised host and internal targets,
\item then exploits the \gls{SSTI} vulnerability to achieve \gls{RCE} on the target web server and stage sensitive system configurations or password hashes into temporary files,
\item subsequently initiates lateral movement by forcing the victim server to transfer these staged files back to the attacker's foothold,
\item tunnels this collected data externally to the designated \gls{C2} server, and
\item finally executes a clean-up exploit to delete all temporary files and logs to obfuscate the intrusion.
\end{inparaenum}
This sequence creates distinct network signatures that are identifiable through holistic analysis but remain difficult to detect when viewed as isolated events.
        
        \newcommand{\dataset}[1]{\texttt{#1}}

\section{Results}
\label{sec:results}

\subsection{Experimental Setup}
\subsubsection{Data Collection and Processing}
\newcommand{\livedataprocessing}{A}
\newcommand{\decoupledprocessing}{B}
\begin{figure}[!t]
    \centering
        \includegraphics[width=0.85\columnwidth]{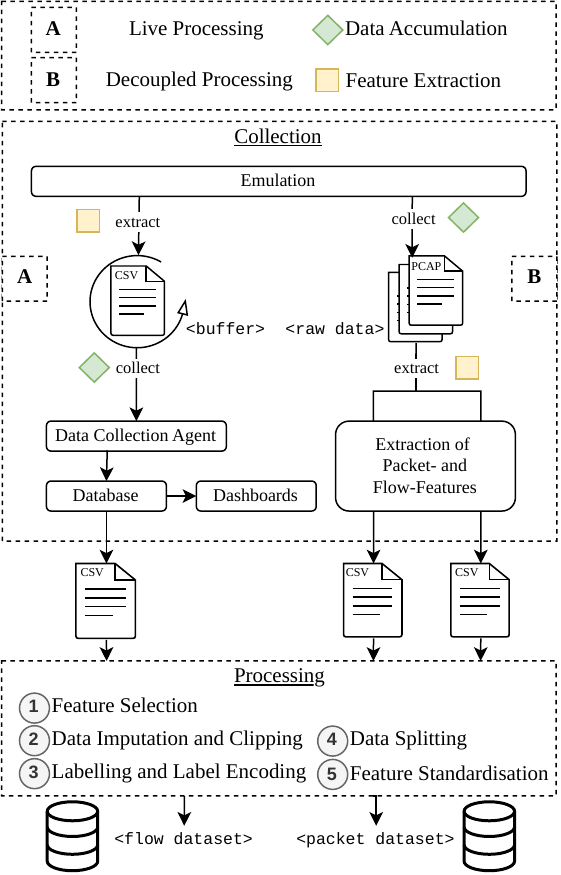}
    \caption{Dual data processing pipeline, supporting live (\livedataprocessing) and decoupled (\decoupledprocessing) data processing.}
    \label{fig:data_pipeline}
    \vspace{-1.0\baselineskip}
\end{figure}

To address the diverse requirements of \gls{ML} algorithms and real-time operational feedback, our methodology employs a dual data pipeline that allows for live (\livedataprocessing) and decoupled data processing (\decoupledprocessing), as illustrated in Fig.~\ref{fig:data_pipeline}. In this figure, the first pipeline (\livedataprocessing) is dedicated to real-time monitoring and visualisation in dashboards, as well as serving as the foundation for \gls{DRL} action-observation loops, necessitating minimal latency. Captured network traffic is immediately processed to extract features into a ring buffer in CSV format. Telegraf, a data acquisition agent, periodically collects and aggregates these features into a time-series InfluxDB database. Grafana is used to organise the data in insightful dashboards, providing real-time analysis capabilities for emulation runs.

The second pipeline (\decoupledprocessing) focusses on deferred batch processing for generating the final synthetic datasets. All captured \acrshort{IP} traffic is collected in batches of PCAP files, which are aggregated and processed post-emulation. This batch approach enables flexible experimentation with different feature sets without the computational overhead of re-running the entire emulation.

The data processing pipeline integrates \emph{tshark} and \emph{NexusFlowMeter} to extract 26 features, including transport metrics, directionality, and payload statistics. To ensure a robust dataset, missing values are addressed via median-based imputation, and outliers are clipped at the $99.99^{th}$ percentile. Ground-truth labels are automatically derived from custom \gls{HTTP} headers at the packet-level and propagated to flows using a worst-case aggregation strategy. Next, the categorical labels are numerically encoded. Finally, the data is partitioned into a stratified 80/20 train-test split and normalized via z-score standardization to facilitate stable model convergence.

\subsubsection{Datasets}
We generated four distinct datasets based on variations of the scenario defined in Section~\ref{sec:methodology}. These variations serve as a testbed for evaluating the robustness and transferability of the trained classifier against different forms of data drift and class imbalance.

For the baseline dataset (\dataset{a}), we selected an emulation duration of 3 hours and a behavioural parametrisation with respect to delays between actions that would result in a higher class-balance, making the dataset more suitable for use in \gls{ML} algorithms. To quantify the transferability of a model trained on this artificially balanced dataset under domain and balance shifts, we generated three additional versions:
\begin{inparaenum}[(1)]
    \item Dataset \dataset{b}: Employs more realistic delays, resulting in naturally sparser network traces and a significantly smaller fraction of malicious flows.
    \item Dataset \dataset{c}: Extends the baseline emulation to 5 hours, resulting in a larger control set.
    \item Dataset \dataset{d}: Introduces a domain-shift barrier for the trained model. This version is heavily dominated by benign \gls{UDP} traffic resulting from unresolved \gls{DNS} queries to the internal \gls{DNS} server in order to send browser telemetry data to an external \acrshort{API}. 
\end{inparaenum}
The resulting protocol and class distributions are listed in Table~\ref{tab:distribution} and discussed in Section~\ref{sec:dataset_analysis}.

\subsubsection{\gls{DNN} Classifier}
The experiments employ a \gls{DNN}-based classification model tailored for multi-class network traffic categorization. A 26-dimensional input vector is processed through four fully connected layers (256, 128, 64, and 32 neurons) utilizing \gls{ReLU} activation and a tapering dropout strategy ($0.3$ to $0.1$) to mitigate overfitting. The five-neuron output layer employs a Softmax activation function, with the model optimized via Sparse Categorical Cross-Entropy loss. Training was conducted for a maximum of 20 epochs using the Adam optimizer with an initial learning rate of $0.0015$, a batch size of 32, and a 0.1 validation split. To ensure stable convergence, the data were shuffled before each epoch, and the regime incorporated an early stopping mechanism (patience of five epochs, restoring optimal weights) and a learning rate scheduler that reduced the rate by a factor of $0.2$ upon validation loss stagnation, down to a minimum of $0.001$.

\subsection{Experimental Results}

\subsubsection{Dataset Analysis}
\label{sec:dataset_analysis}

\begin{figure}[!t]
    \centering
        \includegraphics[width=1.05\columnwidth]{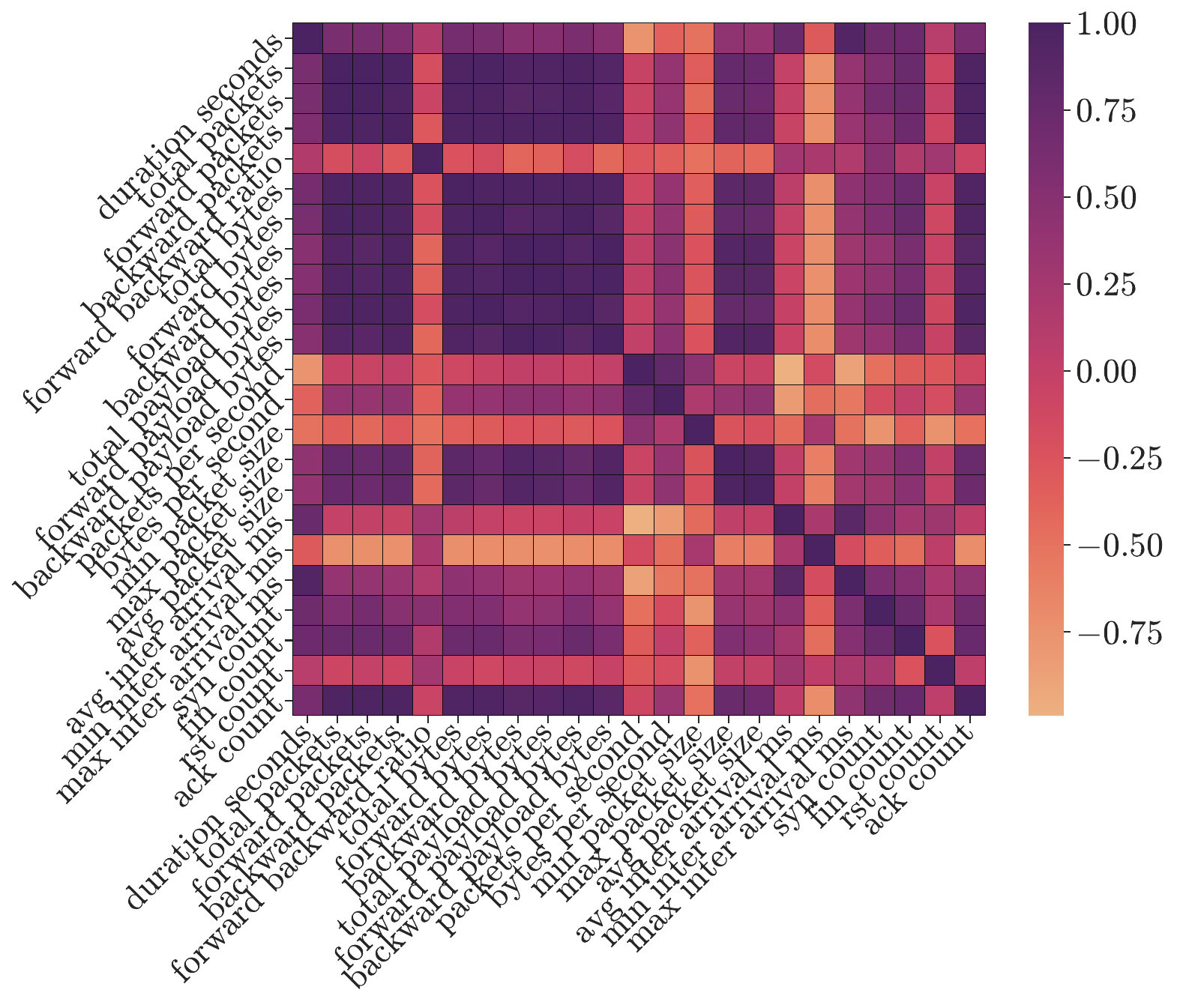}
    \caption{Correlation matrix using Spearman's rank correlation coefficient ($\rho$) (numerical features).}
    \label{fig:correlation_heatmap}
\end{figure}

\begin{table}[b]
\centering
\caption{Protocol and class distribution of each dataset.}
\label{tab:distribution}
\begin{tabular}{lrrrr}
\toprule
Protocol & a & b & c & d \\
\midrule
ARP & 0.103\% & 0.135\% & 0.063\% & 0.013\% \\
ICMP & 0.048\% & 0.042\% & 0.025\% & 0.013\% \\
IGMPv3 & 0.021\% & 0.025\% & 0.012\% & 0.003\% \\
OSPF & 0.027\% & 0.034\% & 0.016\% & 0.003\% \\
TCP & 79.452\% & 81.820\% & 79.887\% & 9.491\% \\
UDP & 20.349\% & 17.943\% & 19.996\% & 90.476\% \\
\midrule %
Class \\
\midrule
Benign & 90.756\% & 98.501\% & 90.653\% & 98.883\% \\
Malicious & 9.244\% & 1.499\% & 9.347\% & 1.117\% \\
Clean-up & 1.661\% & 0.271\% & 1.661\% & 0.203\% \\
Staging & 1.675\% & 0.271\% & 1.686\% & 0.205\% \\
Lateral-movement & 4.118\% & 0.627\% & 4.214\% & 0.500\% \\
Exfiltration & 1.791\% & 0.330\% & 1.786\% & 0.209\% \\
\bottomrule
\end{tabular}
\vspace{-1.0\baselineskip}
\end{table}

The analysis of both class and protocol distributions across the four dataset iterations (Table~\ref{tab:distribution}) reveals significant variance in flow volume and composition. Most notably, dataset \dataset{d} is characterised by \gls{DNS} flooding, which accounts for roughly 216,000 out of 238,000 flows, resulting in 90.48\% \gls{UDP} dominance. In contrast, versions \dataset{a}, \dataset{b}, and \dataset{c} remain characterised by predominantly \gls{TCP} traffic (79.45\%-81.82\%). The baseline \dataset{a} contains a total of 29,100 flows, while \dataset{c} expands this to 48,900 flows, enabling the evaluation of transferability to larger data corpora. Dataset \dataset{a} (and by extension the larger variation \dataset{c}) improves the class-balance from 98.50\%-98.88\% benign traffic (\dataset{d} and \dataset{b}) to 90.76\% (90.65\% for \dataset{c}). Malicious flows are distributed more evenly across their respective sub-classes, with the exception of lateral movement, which encompasses 4.12\% of all flows.

Descriptive statistical analysis of the feature distribution reveals a highly heterogeneous, and heavy-tailed distribution, characteristic of real-world network traffic, where a minority of long-duration, high-volume flows coexist with a vast majority of brief, transactional connections. This disparity is most evident in the profound divergence between central tendency measures. For instance, while the mean flow duration reaches 41.161 seconds, the median (50th percentile) remains a mere 0.047 seconds, indicating that over 75\% of traffic is extremely short-lived ($Q_3=0.131$ seconds). Similar positive skewness is observed across volumetric metrics and inter-arrival times, where rare extreme values substantially inflate the mean relative to the median.

Feature correlation was quantified using Spearman's rank correlation coefficient ($\rho$). The resulting heatmap in Fig.~\ref{fig:correlation_heatmap} reveals significant multicollinearity ($\rho\approx1.0$) within clusters of volumetric metrics, such as total counts and their respective directional components, as well as between mean and maximum packet sizes. In contrast, features such as the average inter-arrival time and transmission rates expectedly exhibit a negative correlation ($\rho\approx-1.0$), reflecting the physical nature that higher throughput typically coincides with lower latencies. Finally, connection state indicators, such as \gls{TCP} flag counts, show near-zero correlation with general flow metrics, confirming their utility as independent variables.

\subsubsection{Model Evaluation}
\label{sec:model_evaluation}

\newcommand{\resultbaseoffset}{-3.5cm}
\newcommand{\resultoffset}{-2.9cm}
\begin{figure}[!t]
    \centering
        \includegraphics[width=0.95\columnwidth]{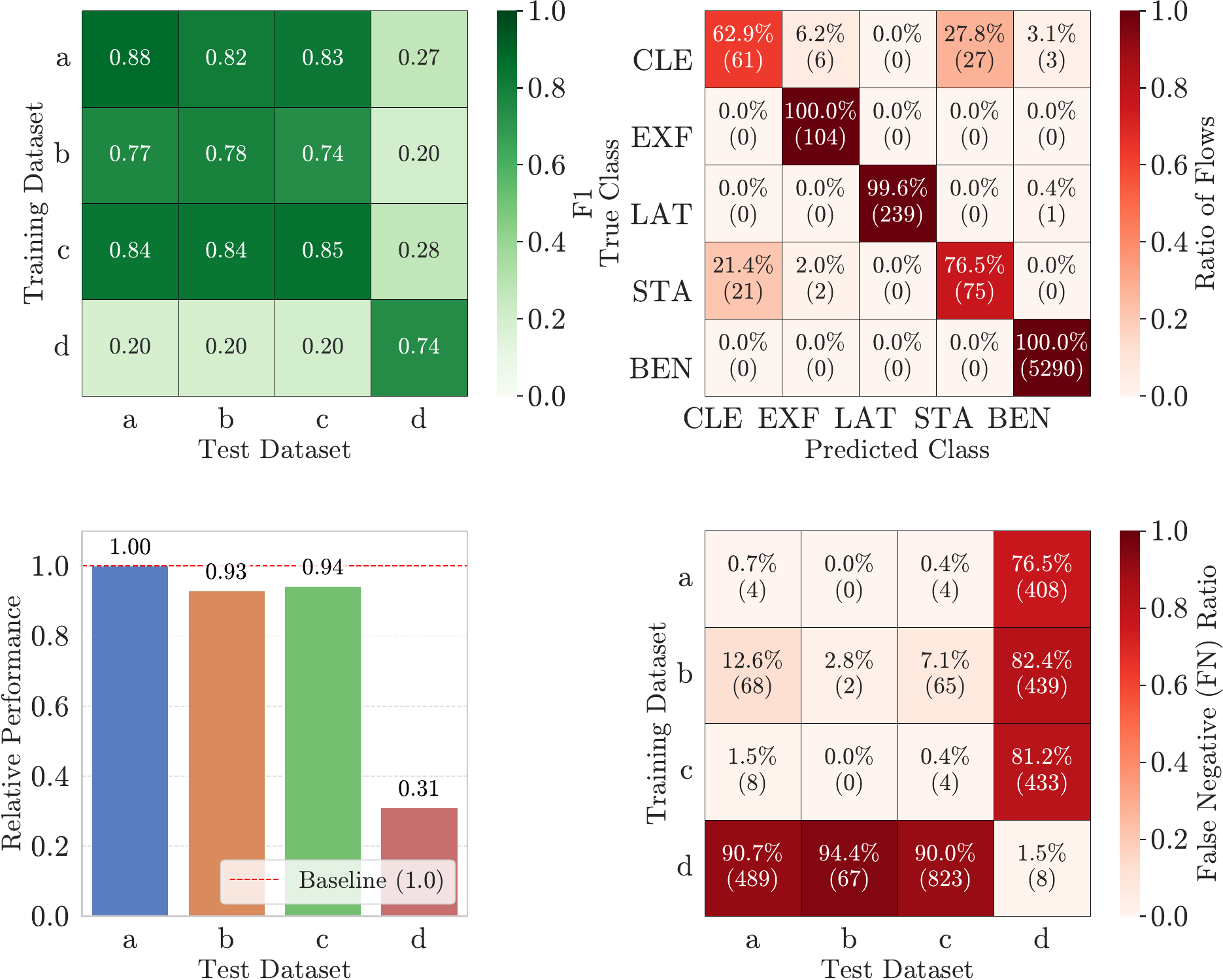}
    \vspace{1.0\baselineskip}
    \caption{Cross-evaluation results: Performance with respect to the F1 score for each model under cross-evaluation (\ref{fig:cross_evaluation}), the confusion matrix of the baseline classifier under in-domain evaluation (\ref{fig:confusion_matrix}), relative F1 score of the baseline classifier when transferred to shifted datasets (\ref{fig:relative_performance}), and the aggregated false negatives under cross-evaluation (\ref{fig:false_negatives}).}
    \label{fig:modelevaluation}
    \vspace{-2.0\baselineskip}

    \begin{subfigure}[t]{0.485\columnwidth}
        \centering
        \vspace{\resultbaseoffset+\resultoffset}
        \caption{Cross Evaluation.}
        \label{fig:cross_evaluation}
    \end{subfigure}%
    \begin{subfigure}[t]{0.485\columnwidth}
        \centering
        \vspace{\resultbaseoffset+\resultoffset}
        \caption{Confusion Matrix.}
        \label{fig:confusion_matrix}
    \end{subfigure}
    \vskip\baselineskip %
    \begin{subfigure}[t]{0.485\columnwidth}
        \centering
        \vspace{\resultbaseoffset}
        \caption{Relative F1 score.}
        \label{fig:relative_performance}
    \end{subfigure}%
    \begin{subfigure}[t]{0.485\columnwidth}
        \centering
        \vspace{\resultbaseoffset}
        \caption{False negative ratio.}
        \label{fig:false_negatives}
    \end{subfigure}
    \vspace{-1.0\baselineskip}
\end{figure}

To assess cross-domain generalization under systematic protocol-distribution shift, we trained a model on each dataset and evaluated its performance on all dataset variants, including the one used for training.
Performance is quantified by the F1 score to assess the classification quality across the five classes `benign', `staging', `lateral-movement', `exfiltration', and `clean-up' with respect to the \glspl{TP}, \glspl{FP}, and \glspl{FN}. 

Fig.~\ref{fig:cross_evaluation} illustrates the F1 score under cross-evaluation for each training-evaluation combination as a heatmap. The diagonal shows high in-domain performance, with F1 scores of 0.88 to 0.85 for the balanced dataset versions \dataset{a} and \dataset{c}, and slightly lower scores of 0.78 to 0.74 for the less balanced datasets \dataset{b} and \dataset{d}, indicating the benefit of utilising the controlled environment to generate artificially balanced datasets for \gls{ML} applications. Fig.~\ref{fig:confusion_matrix} presents the confusion matrix of the baseline classifier as a heatmap. The model achieves perfect recall of 1.0 on benign flows and near perfect recall on the malicious classes `exfiltration' and `lateral-movement' (1.0 and 0.996). For flows that belong to the staging or clean-up stages of the attack chain, the model exhibits noticeably lower recall of 0.765 and 0.629, respectively. Specifically, these two classes are commonly mixed up by the model due to the similarity of their flow statistics, as both comprise a small payload consisting of short system-code, with 21.4\% of the flows associated with the staging phase being falsely attributed to the clean-up stage (conversely, 27.8\% of `clean-up' flows are misclassified as `staging'). 

Fig.~\ref{fig:relative_performance} compares the relative performance with respect to the F1 score of the baseline classifier when evaluated on each dataset variation. Noticeably, performance transfers well to datasets \dataset{b} (more realistic action delays) and \dataset{c} (larger dataset size), with relative scores of 0.93 and 0.94 and corresponding F1 scores of 0.82 and 0.83 respectively (Fig.~\ref{fig:cross_evaluation}). Performance drops significantly under domain shift, where protocol distributions (e.g., ratio of \gls{UDP} to \gls{TCP}) deviate from the training baseline. Specifically, the \gls{DNS} dominated dataset \dataset{d} imposes a domain-shift barrier, with a relative performance drop of 69\% (F1 score of 0.27), showcasing that the vastly different dataset distribution leads to significantly differing decision boundaries learnt by the neural network.

Finally, Fig.~\ref{fig:false_negatives} depicts the aggregate false negatives (instances where malicious flows were predicted as `benign'). The baseline classifier again demonstrates high transferability to the control datasets \dataset{b} and \dataset{c}, even reducing the number of false negative predictions from 0.7\% to 0.0\% for version \dataset{b} compared to a false negative rate of 1.5\% for the in-domain evaluation of the classifier trained on dataset \dataset{b} directly. These results suggest that a classifier can be trained on artificially balanced datasets optimised for \gls{ML} algorithms while maintaining high transferability to more realistic data distributions.
        
        \section{Conclusion}
\label{sec:conclusion}
This study evaluated the efficacy of \gls{ACO} gyms for the development and validation of \gls{ACD} solutions, specifically examining \gls{DNN}-based classifiers trained on synthetic, emulated datasets. While the framework provides a robust baseline, our results reveal a significant ``transferability gap'' when models are subjected to protocol-shifted variations, such as \gls{DNS}/\gls{UDP} flooding. We identify domain shifts as a primary performance barrier, with F1 score dropping to 0.27 when network compositions deviate substantially from training distributions. These findings underscore the necessity for \gls{ACO} frameworks to systematically integrate diverse real-world stressors to ensure the operational reliability of automated defences in critical infrastructure.

To address these limitations, future work will focus on a systematic comparison across a broader range of baseline models beyond \glspl{DNN} to isolate architecture-specific vulnerabilities. Furthermore, we intend to validate our synthetic, emulated datasets against diverse real-world benchmarks to bridge the gap between controlled environments and live production traffic.
Furthermore, we will transition from static threat modelling toward automated attack-chain generation using \gls{AI}-driven agents. Coupled with hybrid architectures, integrating statistical flow analysis, \gls{LLM}-based \gls{DPI}, and host-level telemetry such as system logs and process information, this approach will provide the multi-layered visibility required to secure live production environments against diverse and non-deterministic threats.
  
        \balance
        
        \bibliographystyle{IEEEtran}
        \bibliography{references}
\end{document}